\documentclass[12pt]{article}

\usepackage{multirow}
\usepackage{color}

\usepackage{comment}

\usepackage{times}
\usepackage{bm}

\usepackage{amsmath}
\usepackage{amssymb}
\usepackage{amsfonts}

\usepackage{amssymb}
\usepackage{amsmath,amsthm, mathtools,commath}
\usepackage{graphics, color}

\usepackage{graphicx}
\usepackage{epsfig}
\usepackage{makeidx}
\usepackage[round]{natbib}

\newcommand{\biblist}{\begin{list}{}
{\listparindent 0.0cm \leftmargin 0.50cm \itemindent -0.50 cm
\labelwidth 0 cm \labelsep 0.50 cm
\usecounter{list}}\clubpenalty4000\widowpenalty4000}
\newcommand{\ebiblist}{\end{list}}

\usepackage{latexsym}
\usepackage{amsmath, amssymb, amsfonts, amsthm, bbm}
\usepackage{graphicx}
\usepackage{mathrsfs}

\newtheorem{theorem}{Theorem}

\newtheorem{lemma}{Lemma}

\DeclareMathOperator*{\argmin}{arg\,min}

\newcommand{\bA}{\mathbf{A}}

\newcommand{\bx}{{x}}

\newcommand{\bd}{{d}}
\newcommand{\by}{{y}}

\newcommand{\bz}{\bm{z}}

\newcommand{\bmm}{\bm{m}}

\newcommand{\bI}{\mathbf{I}}

\newcommand{\bC}{\mathbf{C}}
\newcommand{\bK}{\mathbf{K}}

\newcommand{\bS}{\mathbf{S}}

\newcommand{\bone}{\mathbf{1}}

\newcommand{\bvar}{\boldsymbol{\varepsilon}}

\newcommand{\balpha}{\boldsymbol{\alpha}}

\newcommand{\bDelta}{\boldsymbol{\Delta}}

\newcommand{\bPi}{\boldsymbol{\Pi}}

\newcommand{\logit}{\mathrm{logit\,}}

\def\trans{^{\rm T}}
\usepackage{mathabx}
\def\wh{\widehat}
\def\wt{\widetilde}
\newcommand{\Norm}[1]{\left\Vert#1\right\Vert}
\newcommand{\Abs}[1]{\left\vert#1\right\vert}

\usepackage{natbib}

\providecommand{\keywords}[1]{\textbf{Key words:} #1}

\begin{document}

\baselineskip .3in

\title{Statistical Inference after Kernel Ridge Regression Imputation under item nonresponse}

\author{Hengfang Wang \and Jae Kwang Kim}

\date{} 

\maketitle

\begin{abstract}
Imputation is a popular technique for handling missing data. 
We consider a nonparametric approach to imputation using the kernel ridge regression technique and {propose consistent variance estimation}. The proposed variance estimator is based on a linearization approach which employs the entropy method to estimate the density ratio.
The $\sqrt{n}$-consistency of the imputation estimator is established when a Sobolev space is utilized in the kernel ridge regression imputation, which enables us to develop the proposed variance estimator.
Synthetic data experiments are presented to confirm our theory. 
\end{abstract}


%\bibliographystyle{chicago}
%\bibliography{ref}

\newpage

\section{Introduction}

Missing data is a universal problem in statistics.  Ignoring the cases with missing values  can lead to  misleading results \citep{kim2013statistical, little2019statistical}. To avoid the potential problem with missing data, imputation is commonly used.  
 After imputation, the imputed dataset can serve as a complete dataset that has no missing values, which in turn makes results from different analysis methods consistent. However, treating imputed data as if observed and applying the standard estimation procedure may result in misleading inference, leading to  underestimation of the variance  of  imputed point estimators.  As a result, how to make statistical {inferences} with  {imputed point estimators} is an important statistical problem. 
An overview of imputation method can be found in \citet{haziza2009imputation}. 

Multiple imputation,  proposed by \citet{rubin2004multiple}, addresses the uncertainty associated with imputation.  
However,  variance estimation using Rubin's formula  requires  certain conditions \citep{wang1998large,kim2006bias,yang2016note}, which do not necessarily hold in practice. An alternative method  is fractional imputation, originally proposed by \citet{kalton1984some}. The main idea of fractional imputation is to generate multiple imputed values and the corresponding fractional weights. 
%Hot deck imputation is a popular method of imputation  where the imputed values are taken  from  the observed values. In this vein, \citet{fay1996alternative,kim2004fractional,fuller2005hot,durrant2005imputation,durrant2006using} discussed fractional hot deck imputation.
In particular,
\citet{kim2011parametric} and \citet{kim2014fractional} employ fully parametric approach to handling nonresponse items with fractional imputation. However, such parametric fractional imputation relies heavily on the parametric model assumptions.  To mitigate the effects of parametric model assumption, empirical likelihood \citep{owen2001empirical,qin1994empirical} as a semiparametric approach was considered. In particular,  \citet{wang2009empirical} employed the kernel smoothing approach to do empirical likelihood inference with missing values. %\citet{chen2017semiparametric} extended \citet{muller2009estimating}'s work to develop fractional imputation with the first moment  assumption only.   
\citet{cheng1994nonparametric} utilized the kernel-based nonparametric regression approach to do the imputation and established the $\sqrt{n}$-consistency of the imputed estimator.

Kernel ridge regression \citep{friedman2001elements,shawe2004kernel} is a popular data-driven approach which can alleviate the effect of model assumption. {By using} a regularized \textit{M}-estimator {in reproducing} kernel Hilbert space (RKHS), kernel ridge regression can capture the model with {complex reproducing kernel Hilbert space} while a regularized term makes the original infinite dimensional estimation problem viable \citep{wahba1990spline}. \citet{geer2000empirical,mendelson2002geometric,zhang2005learning,koltchinskii2006local,steinwart2009optimal} studied the error bounds for the estimates of kernel ridge regression method. 
%Recently, \citet{zhang2013divide} employed truncation analysis to estimate the error bound in a distributed fashion. \citet{yang2017randomized} considered randomized sketches for KRR and studied projection dimension which can preserve minimax optimal approximations for KRR.

%\subsection{Our Work}

In this paper, we apply kernel ridge regression as a nonparametric imputation method and propose a consistent variance estimator for the corresponding imputation estimator under missing at random framework. Because the kernel ridge regression is a general tool for nonparametric regression with flexible assumptions, the proposed imputation method is practically useful. Variance estimation after the kernel ridge regression imputation is a challenging but important problem.   
To the best of our knowledge, this is the first paper which considers kernel ridge regression technique and discusses its variance estimation in the imputation framework.  Specifically, we first prove $\sqrt{n}$-consistency of the kernel ridge regression imputation estimator and obtain influence function for linearization. After that, we 
employ the maximum entropy method \citep{nguyen2010} for density ratio estimation to get a valid estimate of the inverse of the propensity scores. The consistency of our variance estimator can then  be established.

The paper is organized as follows. In Section 2, the basic setup and  the proposed method are introduced. In Section 3, main theory is established. We also introduce a novel  nonparametric estimator of the propensity score function.  Results from two limited simulation studies are presented in Section 4. An illustration of the proposed method to a real data example is presented in Section 5. Some concluding remarks are made in Section 6.

\section{Proposed Method}

Consider the problem of estimating $\theta=\mathbb{E}(Y)$ from an independent and identically distributed (IID) sample  $\{(\bx_i, y_i), i=1, \cdots, n\}$ of random {vector} $(X,Y)$. Instead of always {observing  $y_i$}, suppose that we observe $y_i$ only if $\delta_i=1$, where $\delta_i$ is the response indicator function of unit $i$ taking values on $\{0,1\}$.  The auxiliary variable $\bx_i$ are always observed. 
We assume that the response mechanism is missing at random (MAR)   in the sense of   \cite{rubin1976}. 
%$$ \delta \perp Y \mid \bx . $$

Under MAR, we can develop a nonparametric estimator $\wh{m} (\bx)$ of $m(\bx)=\mathbb{E}( Y \mid \bx)$ and construct the following imputation estimator: 
\begin{equation} 
\wh{\theta}_I = \frac{1}{n} \sum_{i=1}^n \left\{ \delta_i y_i + (1-\delta_i )\wh{m} (\bx_i) \right\}. 
\label{1} 
\end{equation} 
If $\wh{m} (\bx)$ is constructed by the kernel-based nonparametric regression method, we can express 
\begin{equation} 
 \wh{m} (\bx) = \frac{ \sum_{i=1}^n \delta_i K_h( \bx_i, \bx ) y_i}{ \sum_{i=1}^n \delta_i K_h( \bx_i, \bx )} 
\label{2}
\end{equation} 
where $K_h (\cdot)$ is the kernel function with bandwidth $h$. Under some suitable choice of the bandwidth $h$, \citet{cheng1994nonparametric} first established the $\sqrt{n}$-consistency of the imputation estimator (\ref{1}) with nonparametric  function in (\ref{2}).  However, the kernel-based regression imputation in (\ref{2}) is applicable only when the dimension of $x$ is small.

%\textcolor{red}{Suppose we can express 
%$$ \hat{\theta}_I = \frac{1}{n} \sum_{i=1}^n \hat{m}(\bx_i) = \frac{1}{n} \sum_{i =1}^n\delta_i  \hat{g} (\bx_i) y_i , $$
%then 
% $\hat{g}(\bx)$ is a nonparametric estimator of $1/\pi(\bx)$. Thus, the linearization form of $\hat{\theta}_I$ is 
% $$ \tilde{\theta}_I = \frac{1}{n} \sum_{i=1}^n \{ \hat{m}_i+ \delta_i \hat{g}( \bx_i) (y_i - \hat{m}_i  ) \}$$
% where $\hat{m}_i = \hat{m} (\bx_i)$. The $\hat{g}( \bx)$ satisfies 
% $$ \sum_{i=1}^n \hat{m}_i = \sum_{i=1}^n\delta_i  \hat{g} (\bx_i) \hat{m}_i . $$
%}
%

In this paper, we extend the work of \citet{cheng1994nonparametric} by considering a more general type of the nonparametric imputation, called kernel ridge regression (KRR) imputation. The KKR technique can be understood using the reproducing kernel Hilbert space (RKHS) theory \citep{aronszajn1950theory} and can be described as 
\begin{equation} 
\wh{m} = \argmin_{m\in \mathcal{H}} \left[   \sum_{i=1}^{n} \delta_{i}\left\{ y_{i} - m(\bx_{i}) \right\}^{2} + \lambda \Norm{m}_{\mathcal{H}}^{2}   \right],
\label{3} 
\end{equation} 
where $ \norm{m}_{\mathcal{H}}^{2}  $ is the norm of $m$ in the Hilbert space $\mathcal{H}$. Here, the inner product $\langle \cdot, \cdot \rangle_{\mathcal{H}}$ is induced by {such} a kernel function, i.e., 
\begin{align}
\langle f, K(\cdot, \bx) \rangle_{\mathcal{H}} = f(\bx), \forall \bx \in \mathcal{X},  f\in \mathcal{H},
\end{align}
namely, the reproducing property of $\mathcal{H}$. Naturally, this  reproducing property implies the $\mathcal{H}$ norm of $f$: $\norm{f}_{\mathcal{H}} = \langle f, f  \rangle_{\mathcal{H}}^{1/2}$.

One 
canonical example of such a space is the Sobolev space. Specifically, assuming that  the domain of such functional space is $[0,1]$,
 the Sobolev space of order $l$ can be denoted as 
\begin{eqnarray}
	\mathcal{W}_{2}^{l} &=& \left\{ f:[0,1] \rightarrow \mathbb{R} |
	 f, f^{(1)}, \dots, f^{(l-1)} \mbox{ are absolute continuous and } f^{(l)} \in L^{2}[0,1]    \right\}. \notag 
\end{eqnarray}
One possible norm for this space can be
\begin{eqnarray}
	\norm{f}_{\mathcal{W}_{2}^{l}}^{2} = \sum_{q = 0}^{l-1}\left\{   \int_{0}^{1}f^{(q)}(t)dt      \right\}^{2} +
	 \int_{0}^{1}\left\{f^{(l)}(t) \right\}^{2}dt . \notag 
\end{eqnarray}
%Readers can refer to \cite{wahba1990spline} for a thorough treatment of the RKHS technique. 
In this section, we employ the Sobolev space of second order as the approximation space. 
For Sobolev space of order $\ell$, we have the kernel function
\begin{align}
K(x,y) = \sum_{q = 0}^{\ell-1}k_{q}(x)k_{q}(y) + k_{\ell}(x)k_{\ell}(y)
 + (-1)^{\ell}   k_{2\ell}(|x-y|),\notag 
\end{align}
where $k_{q}(x) = (q!)^{-1}B_{q}(x)$ and $B_{q}(\cdot)$ is the Bernoulli polynomial of order $q$.

By the representer theorem for RKHS \citep{wahba1990spline}, the estimate  in (\ref{3})  lies in the linear span of $\{K(\cdot, \bx_{i}), i = 1,\ldots, n\}$.
Specifically, we have 
\begin{align}\label{KRR}
\wh{m}(\cdot) = \sum_{i=1}^{n}\wh{\alpha}_{i,\lambda}K(\cdot, \bx_{i}),
\end{align}
where 
\begin{align}
\wh{\balpha}_{\lambda} = \left(\bDelta_{n}\bK +  \lambda\bI_{n}\right)^{-1}\bDelta_n \by,\notag
\end{align}
$\bDelta_{n} = \mbox{diag}(\delta_{1}, \ldots, \delta_{n})$, $\bK = (K(\bx_{i}, \bx_{j}))_{ij} $,  $\by = (y_{1},\ldots, y_{n})\trans$ and $\bI_{n}$ is the $n\times n$ identity matrix.

The tuning parameter $\lambda$ is selected via generalized cross-validation (GCV) in KRR, where the GCV criterion for $\lambda$ is
\begin{align}\label{GCV}
 \mbox{GCV}(\lambda) = \frac{n^{-1}   \Norm{ \left\{\bDelta_{n} - \bA(\lambda)\right\}\by }_{2}^{2}   }{n^{-1}  \mbox{Trace}(\bDelta_{n} - \bA(\lambda) )  },
\end{align}
and $\bA(\lambda) = \bDelta_{n}\bK  ( \bDelta_{n} \bK + \lambda \bI_{n}  )^{-1} \bDelta_{n}  $. The value of $\lambda$ minimizing the GCV is used for the selected tuning parameter.

Using the KRR imputation in (\ref{3}), we aim to establish the following two goals:
\begin{enumerate}
	\item Find the sufficient conditions for the   $\sqrt{n}$-consistency of the imputation estimator $\wh{\theta}_I$ using (\ref{KRR}) and give a formal proof. 
	\item Find a linearization variance formula for the imputation estimator $\wh{\theta}_I$ using the KRR imputation. 
\end{enumerate}
The first part is formally presented in Theorem 1 in Section 3. For the second part, we employ the density ratio estimation method of \cite{nguyen2010} 
to get a consistent estimator of  $\omega (\bx) = \{\pi (\bx)\}^{-1}$ in the linearized version of $\hat{\theta}_I$. 
\begin{comment}
By Theorem 1, we use the following estimator to estimate the variance of $\wh{\theta}_I$ in (\ref{2}): 
\begin{align}\label{variance estimation}
\wh{\mbox{V}} (\wh{\theta}_{I}) = \frac{1}{n(n-1)}\sum_{i=1}^{n}\left( \hat{\eta}_i - \bar{\eta} \right)^2 
\end{align}
where 
$$  \hat{\eta}_i =  \wh{m}(\bx_{i}) + \delta_{i} \wh{\omega}_{i} \left\{ y_{i} - \wh{m}(\bx_{i})\right\},
$$
and $ \wh{\omega}_{i} $ is a consistent estimator of $\omega (\bx) = \{\pi (\bx)\}^{-1}$. 
\end{comment}

\section{Main Theory}

Before we develop our main theory, we first introduce Mercer's theorem. 
\begin{lemma}[Mercer's theorem]\label{Mercer}
Given a continuous, symmetric, positive definite kernel function $K: \mathcal{X} \times \mathcal{X} \mapsto \mathbb{R}$. For $\bx, \bz \in \mathcal{X}$, under some regularity conditions, Mercer's theorem characterizes $K$ by the following expansion
\begin{align}
 K(\bx, \bz) = \sum_{j=1}^{\infty}\lambda_{j}\phi_{j}(\bx) \phi_{j}(\bz),\notag
\end{align}
where $\lambda_{1} \geq \lambda_{2} \geq \ldots \geq 0$ are a non-negative sequence of eigenvalues and $\{\phi_{j} \}_{j=1}^{\infty}$ is an orthonormal basis for $L^{2}(\mathbb{P})$.
\end{lemma}

To develop our theory, we  make the following  assumptions.
 \begin{description}
 
 \item {[A1]}
 \label{A1}
 	For some $k \geq 2$, there is a constant $\rho < \infty$ such that $E[ \phi_{j}(X)^{2k} ] \leq \rho^{2k}$ for all 
 	$j \in \mathbb{N}$, where $\{\phi_{j}\}_{j=1}^{\infty}$ are orthonormal basis by expansion from Mercer's theorem.

 \item {[A2]}
 \label{A2}
 	The function $m \in \mathcal{H}$, and for $\bx \in \mathcal{X}$, we have $E[\left\{ Y -  m(\bx)\right\}^{2} ] \leq \sigma^{2}$, {for some  $\sigma^{2} < \infty$.}

\item {[A3]}
\label{A3}
     The propensity score $\pi(\cdot)$ is uniformly bounded away from zero. In particular, there exists a positive constant $c > 0$ such that 
     	$\pi(\bx_{i}) \geq c$, for $i = 1, \ldots, n$.

\item {[A4]}
 \label{A4}
     The ratio $d/\ell < 2$ for $d$-dimensional Sobolev space of order $\ell$, where $d$ is the dimension of covariate $\bx$.
\end{description}

The first assumption is a technical assumption which controls the tail behavior of $\{\phi_{j}\}_{j=1}^{\infty}$. Assumption 2 indicates that the noises have bounded variance. Assumption 1 and Assumption 2 together aim to control the error bound of the kernel ridge regression estimate $\wh{m}$. {Furthermore}, Assumption 3 means that the support for the respondents should be the same as  the original sample support. Assumption 3 is a standard assumption for missing data analysis. Assumption 4 is a technical assumption for entropy analysis. Intuitively, when the dimension is large, the Sobolev space should be large enough to capture the true model.

\begin{theorem}\label{main theorem}
Suppose Assumption $1 \sim 4$ hold for a Sobolev kernel of order $\ell$, $\lambda \asymp   n^{1-\ell}$, we have
\begin{align}\label{rate}
 \sqrt{n}(\wh{\theta}_{I} - \wt{\theta}_{I} ) = o_{p}(1), 
\end{align}
where 
\begin{align}\label{tilde_theta}
\wt{\theta}_{I} &= \frac{1}{n}\sum_{i=1}^{n} \left[ m(\bx_{i}) + \delta_{i} \frac{1}{\pi(\bx_{i})}  \left\{ y_{i} - m(\bx_{i})\right\} 
   \right] 
\end{align}
and 
\textcolor{blue}{}
$$\sqrt{n} \left( \tilde{\theta}_I - \theta \right) \stackrel{\mathcal{L}}{\longrightarrow}  N(0, \sigma^2 ) ,
$$
 with 
$$ \sigma^2 = V\{ E( Y \mid \bx) \} + E\{ V( Y \mid \bx)/\pi( \bx)   \} . $$ 

%Let $\omega_{i}^{\star} = \pi(\bx_{i})^{-1}$. In particular,  $\pmb{\omega}^{\star} = \{\omega_{i}^{\star}: \delta_{i} = 1\}$  can be estimated by\cite{wong2018kernel}.
\end{theorem}

Theorem \ref{main theorem} guarantees the asymptotic equivalence of $\wh{\theta}_{I}$ and $\wt{\theta}_{I}$ in \eqref{tilde_theta}. Specifically, the reference distribution is a combination of an outcome model and a propensity score model for sampling mechanism. The variance of $\wt{\theta}_I$ achieves the semiparametric lower bound of \citet{robins94}. 
%{Additionally, \eqref{tilde_theta} suggests a linearization form of variance estimation of $\wh{\theta}_{I}$. To estimate $\omega_{i}^{\star} = \pi(\bx_{i})^{-1}$, we can use employ the covariate balancing idea of \citet{wong2018kernel} to obtain $\hat{\pmb{\omega}}$ as provided in \eqref{WC_Opt}. }  
The proof of Theorem \ref{main theorem} is presented in the Appendix.
%\hf{Here, in some sense, I think we do not have a valid variance estimator for \citet{wong2018kernel}. In particular, they have uncertainty of $\wh{\omega}_{i}$. In our case, we only consider the variance estimator for $\wh{\theta}_{I}$, which only involves $\wh{m}$.}

The linearization formula in (\ref{tilde_theta}) can be used for  variance estimation. The idea is to estimate the influence function 
$\eta_i = m(\bx_{i}) + \delta_{i} \{ \pi(\bx_{i})\}^{-1}  \left\{ y_{i} - m(\bx_{i}) \right\} $ and apply the standard variance estimator using $\hat{\eta}_i$. To estimate $\eta_i$, we need an estimator of $\pi(x)$. We propose a version of KRR method to estimate $\omega(x) = \{ \pi(x) \}^{-1}$ directly.  
In order to estimate $\omega(x) = \{ \pi(x) \}^{-1}$, we wish to develop a {KRR} version of estimating $\omega (x)$. To do this, first define 
\begin{equation} 
 g( x) = \frac{ f(x \mid  \delta =0 ) }{ f( x \mid  \delta = 1 ) },
\label{dr2}
\end{equation} 
and, by Bayes theorem, we have  
$$ \omega(x)= \frac{1}{ \pi(x) }  = 1+ \frac{n_0}{n_1}  g(x).
$$
Thus, to estimate $\omega(x)$, we have only to estimate the density ration function $g(x)$ in (\ref{dr2}). 
Now, to estimate $g(x)$ nonparametrically, we use the idea of \cite{nguyen2010} for  the KRR approach to density ratio estimation. 

%in order to  estimate $g( x)$ and obtain a nonparametric estimator of $\omega(x)$. 

%\jk{Some details on the estimation of $g(x)$ should be placed here } 

To explain the KRR estimation of $g(x)$, note that $g(x)$ can be understood as the  maximizer of 
 \begin{align} 
     Q (g) = \int \log \left( g \right) f( \bx \mid \delta =0)  d \mu(\bx)   -   \int g (x) f( \bx \mid \delta = 1)   d \mu(x)    \label{qq} 
    \end{align} 
    with constraint 
    $$  \int g (x) f( \bx \mid \delta = 1)   d \mu(x) = 1 . 
    $$ 
    The sample version objective function is 
 \begin{equation} 
\hat{Q}( g) =  \frac{1}{n_0} \sum_{i=1}^n \mathbb{I} ( \delta_i=0) \log \{ g(\bx_i)  \} - \frac{1}{n_1} \sum_{i=1}^n \mathbb{I} ( \delta_i=1) g (\bx_i) 
\label{qq2} 
\end{equation}  
where $n_k = \sum_{i=1}^n \mathbb{I} (\delta_i = k ) $. The maximizer of $\hat{Q} (g)$ is an M-estimator of the density ratio function $g$.

%Define 
%$h (\bx)= \log \{ g(\bx)\}$. The loss function  $L ( \cdot)$  derived from the optimization problem in (\ref{qq2}) can be written as 
%\begin{equation*} 
% L( \delta, h(\bx)  ) = \frac{1}{n_0}  \mathbb{I} (\delta=0) h(\bx) - \frac{1}{n_1} \mathbb{I} (\delta=1) \exp \{ h(\bx) \}. 
%\end{equation*} 

Further, define 
$h (\bx)= \log \{ g(\bx)\}$.
 The loss function 
 $L ( \cdot)$  derived from the optimization problem in (\ref{qq2}) can be written as 
\begin{equation*} 
 L( \delta, h(\bx)  ) = \frac{1}{n_1} \mathbb{I} (\delta=1) \exp \{ h(\bx) \} - \frac{1}{n_0}  \mathbb{I} (\delta=0) h(\bx).
\end{equation*} 
%such that  $$ \hat{Q} ( r_k) = \sum_{i=1}^N L( y_i, h_k (x_i) ) . $$

In our  problem, we wish to find $h$ that minimizes 
\begin{equation}
\sum_{i=1}^n L( \delta_i, \alpha_{0} + h(x_i) ) + \tau \left\| h \right\|_{\mathcal{H} }^{2} \label{18} 
\end{equation} 
over $\alpha_0 \in \mathbb{R}$ and $h \in \mathcal{H} $, where $L ( \cdot) $ is the loss function  derived from the optimization problem in (\ref{qq}) using maximum entropy. 
%That is, 

Hence, using the representer theorem again, the solution to (\ref{18}) can be obtained as 
\begin{equation} \label{entropy_method}
\min_{ \alpha \in \mathbb{R}^n  } \left\{ \sum_{i=1}^n L( \delta_i, \alpha_0 + \sum_{j=1}^n   \alpha_j K( x_i, x_j )) + \tau \balpha' \mathbf{K}  \balpha  \right\} 
\end{equation}
and $\alpha_0$ is a normalizing constant satisfying 
$$ n_1 = \sum_{i=1}^n  \mathbb{I} (\delta_i = 1) \exp \{ \alpha_0 + \sum_{j=1}^n  \hat{\alpha}_j K( x_i, x_j ) \} . $$
 Thus, we use 
\begin{equation}
 \hat{g} (x) = \exp \{ \hat{\alpha}_0 + \sum_{j=1}^n  \hat{\alpha}_j K( x, x_j)  \}
 \label{final} 
 \end{equation} 
 as a nonparametric approximation of the density ratio function $g( x)$. Also, 
 \begin{equation} 
 \hat{\omega} (x)= 1+ \frac{n_0}{n_1} \hat{g} (x)
 \label{final2}
 \end{equation} 
 is the nonparametric approximation of $\omega(x) = \{ \pi(x) \}^{-1}$. Note that $\tau$ is the tuning parameter that determines the model complexity of $g(x)$. The tuning parameter selection is discussed in Appendix B.

Therefore, we can use 
\begin{equation} 
\wh{\mbox{V}}  = \frac{1}{n} \frac{1}{n-1} \sum_{i=1}^n \left( \hat{\eta}_i - \bar{\eta}_n \right)^2  
\label{varh}
\end{equation} 
as a variance estimator of $\hat{\theta}_I$, where 
\begin{equation} 
\hat{\eta}_i = \wh{m}(\bx_{i}) + \delta_{i} \wh{\omega}_{i} (x_i) \left\{ y_{i} - \wh{m}(\bx_{i})\right\}
\end{equation} 
and $\bar{\eta}_n = n^{-1} \sum_{i=1}^n \hat{\eta}_i$.

\section{Simulation Study}

%\jk{We need to update our simulation result because we use a different function of $\hat{\omega}(x)$ for variance estimation. } 

%\hf{Simulation results updated.}
\subsection{Simulation study one} 

To evaluate the performance of the proposed imputation method and its variance estimator, we conduct two simulation studies. In the first simulation study, a continuous study variable is considered with three different data generating models.   In the three models, we keep the response rate around $70\%$ and $\mbox{Var}(Y) \approx 10$. Also, $\bx_{i} =  (x_{i1}, x_{i2}, x_{i3}, x_{i4})\trans$ are generated independently {element-wise} from the {uniform} distribution on the support $(1,3)$. In the first model (Model A), we use a linear regression model 
%The responses for linear case (Model A) are generated by 
\begin{align}
y_{i} =& 3 + 2.5x_{i1} + 2.75 x_{i2} +  2.5 x_{i3} + 2.25 x_{i4} + \sigma\epsilon_{i},\notag
\end{align}
to obtain $y_i$, 
where $\{\epsilon_{i}\}_{i=1}^{n}$ are generated from standard normal distribution and $\sigma = \sqrt{3}$. In the second model  (Model B), we use  \begin{align}
y_{i} =& 3 + (1/35)x_{i1}^{2}x_{i2}^{3}x_{i3} +  0.1x_{i4} + \sigma\epsilon_{i} \notag
\end{align}
to generate data with a {nonlinear} structure. The third model  (Model C) for generating the study variable is 
\begin{align}
y_{i} =& 3 + (1/180)x_{i1}^{2}x_{i2}^{3}x_{i3}x_{i4}^{2} + \sigma\epsilon_{i}.\notag
\end{align}

In addition to $\{(x_i\trans, y_i)\trans, i = 1, \ldots, n\}$, the response indicator variable $\delta$'s are independently generated from the {Bernoulli} distribution with probability $\logit(\bx_i' \beta + 2.5)$, where  ${\beta} = (-1, 0.5, -0.25, -0.1)\trans$ and $\mbox{logit}(p) = \log\{p / (1-p)\}$. We considered three sample sizes $n = 200$, $n = 500$ and $n = 1,000$ with 1,000 Monte Carlo replications. The reproducing kernel Hilbert space we employed is the second-order Sobolev space. 

We also compare three imputation methods: kernel ridge regression (KRR), B-spline, linear regression (Linear). We compute the Monte Carlo biases, variance, and the mean squared {errors} of the imputation estimators for each case. 
The corresponding results are presented in   Table \ref{MAR, linear, comparison}.  
%In Table \ref{MAR, linear, comparison}, the performance of the three imputation estimators are presented.   

\begin{table}[!ht]
\centering
\caption{Biases, Variances and Mean Squared Errors (MSEs) of three imputation estimators for continuous responses}\label{MAR, linear, comparison}
{
\begin{tabular}{cccccc}
  \hline
Model & Sample Size & Criteria & KRR & B-spline & Linear  \\ 
  \hline
 \multirow{9}{*}{A}& \multirow{3}{*}{$200$} & Bias & -0.0577 & 0.0027 & 0.0023 \\ 
 & & Var & 0.0724 & 0.0679 & 0.0682 \\ 
 & & MSE & 0.0757 & 0.0679 & 0.0682 \\ 
 \cline{3-6}
 &\multirow{3}{*}{$500$} & Bias & -0.0358 & 0.0038 & 0.0038 \\ 
 & & Var & 0.0275 & 0.0263 & 0.0263 \\ 
 & & MSE & 0.0288 & 0.0263 & 0.0263 \\
 \cline{3-6} 
 &\multirow{3}{*}{$1000$} & Bias & -0.0292 & 0.0002 & 0.0002 \\ 
 & & Var & 0.0132 & 0.0128 & 0.0129 \\ 
 & & MSE & 0.0141 & 0.0128 & 0.0129 \\
   \hline
\multirow{9}{*}{B} &  \multirow{3}{*}{$200$} & Bias & -0.0188 & 0.0493 & 0.0372 \\ 
 & & Var & 0.0644 & 0.0674 & 0.0666 \\ 
 & & MSE & 0.0648 & 0.0698 & 0.0680 \\ 
  \cline{3-6} 
 &\multirow{3}{*}{$500$} & Bias & -0.0136 & 0.0463 & 0.0356 \\ 
 & & Var & 0.0261 & 0.0275 & 0.0272 \\ 
 & & MSE & 0.0263 & 0.0296 & 0.0285 \\ 
  \cline{3-6} 
 &\multirow{3}{*}{$1000$} & Bias & -0.0122 & 0.0426 & 0.0313 \\ 
 & & Var & 0.0121 & 0.0129 & 0.0129 \\ 
 & & MSE & 0.0123 & 0.0147 & 0.0139 \\ 
\hline
  \multirow{9}{*}{C} &  \multirow{3}{*}{$200$}& Bias & -0.0223 & 0.0384 & 0.0283 \\ 
 & & Var & 0.0748 & 0.0811 & 0.0792 \\ 
 & & MSE & 0.0753 & 0.0825 & 0.0800 \\ 
 \cline{3-6}
 &\multirow{3}{*}{$500$} & Bias & -0.0141 & 0.0369 & 0.0287 \\ 
 & & Var & 0.0281 & 0.0307 & 0.0301 \\ 
 & & MSE & 0.0283 & 0.0320 & 0.0309 \\ 
 \cline{3-6}
 &\multirow{3}{*}{$1000$} & Bias & -0.0142 & 0.0310 & 0.0221 \\ 
 & & Var & 0.0124 & 0.0138 & 0.0136 \\ 
 & & MSE & 0.0126 & 0.0148 & 0.0141 \\ 
 \hline
\end{tabular}
}
\end{table}

The simulation results in  Table \ref{MAR, linear, comparison} shows that  the three methods show similar results under the linear model (Model A), but   kernel ridge regression imputation shows the best performance in terms of {the mean square errors} under the nonlinear models (Models B and C). Linear regression imputation still provides unbiased estimates, because the residual terms in the linear regression model are  approximately unbiased to zero. However, use of linear regression model for imputation leads to efficiency loss because it is not the best model.

In addition, we have computed the proposed variance estimator under kernel ridge regression imputation. 
%The behavior of variance estimation for the imputation estimator is presented in Table \ref{MAR, linear, KRR}. 
In Table \ref{MAR, linear, KRR}, the relative biases of the proposed variance estimator  and the coverage rates of two interval estimators under $90\%$ and $95\%$ nominal coverage rates are presented. 
The relative bias of the variance estimator decreases as the sample size increases, which confirms  the validity of the proposed variance estimator. Furthermore, the interval estimators show good performances in terms of the coverage rates. 

%calculated by the variance estimation based on our method are well approximated.

\begin{table}[!ht]
\centering
\caption{Relative biases (R.B.)  of the proposed variance estimator, coverage rates (C.R.) of the $90\%$ and $95\%$ confidence intervals for imputed estimators under kernel ridge regression imputation for continuous responses}\label{MAR, linear, KRR}
\begin{tabular}{ccccc}
  \hline
\multirow{2}{*}{Model} & \multirow{2}{*}{Criteria}  & \multicolumn{3}{c}{Sample Size} \\
  \cline{3-5} 
    &  &   200 & 500 & 1000 \\ 
  \hline
 \multirow{3}{*}{A} & R.B. & -0.1050 & -0.0643 & -0.0315 \\ 
 &  C.R. (90\%) & 87.5\% & 89.6\% & 89.9\% \\ 
 &  C.R. (95\%)  & 94.0\% & 94.7\% & 94.9\% \\ 
 \hline
\multirow{3}{*}{B} & R.B. &-0.1016 & -0.1086 & -0.0276 \\ 
 & C.R. (90\%) & 87.6\% & 87.0\% & 89.2\% \\ 
 & C.R. (95\%)  & 92.6\% & 93.3\% & 94.8\% \\ 
   \hline
 \multirow{3}{*}{C} &  R.B. & -0.1934 & -0.1310 & -0.0054 \\
 &  C.R. (90\%)  & 85.0\% & 86.2\% & 90.4\% \\ 
 &  C.R. (95\%)  & 91.4\% & 93.4\% & 94.6\% \\ 
 \hline
\end{tabular}
\end{table}

\subsection{Simulation study two} 

The second simulation study is similar to the first simulation study except that the study variable $Y$ is binary. We use the same simulation setup for generating $\bx_i = (x_{1i}, x_{2i}, x_{3i}, x_{4i})$ and $\delta_i$ as the first simulation study. We consider three models for generating $Y$ 
\begin{equation}
y_{i} \sim \mbox{Bernoulli}(p_{i}), \label{model2} 
\end{equation}
where $p_{i}$ is chosen differently  for each model. 
For model D, we have
\begin{align}
\mbox{logit}(p_{i}) =  0.5 + (1/35)x_{i1}^{2}x_{i2}^{3}x_{i3} +  0.1x_{i4}.\notag
\end{align}
The responses for Model E are generated by (\ref{model2}) with 
\begin{align}
\mbox{logit}(p_{i}) =  0.5 + (1/180)x_{i1}^{2}x_{i2}^{3}x_{i3}x_{i4}^{2}.\notag
\end{align}
The responses for Model F are generated by (\ref{model2}) with 
\begin{align}
\mbox{logit}(p_{i}) =  0.5 + 0.15x_{i1}x_{i2}x_{i3}^{2} +  0.4x_{i2}x_{i3}.\notag
\end{align}

For each model, we consider three imputation estimators: kernel ridge regression (KRR), B-spline, linear regression (Linear). We compute the Monte Carlo biases, variance, and the mean squared errors of the imputation estimators for each case.  {The comparison of the simulation 
results for different estimators  are} presented in Table \ref{MAR, nonlinear, comparison, DEF}. In addition, the relative biases and the coverage rates of the interval estimators  are presented in Table \ref{MAR, binary, KRR}. The simulation results in Table \ref{MAR, binary, KRR} show that the relative biases of the  variance estimators are negligible  and the coverage rates of the interval estimators are close to the nominal levels.

\begin{table}[!ht]
\centering
\caption{Biases, Variances and Mean Squared Errors (MSEs) of three imputation estimators for binary responses}\label{MAR, nonlinear, comparison, DEF}
{
\begin{tabular}{cccccc}
  \hline

Model & Sample Size & Criterion & KRR & B-spline & Linear  \\ 
  \hline
 \multirow{9}{*}{D} &  \multirow{3}{*}{$200$}& Bias & 0.00028 & 0.00007 & 0.00009 \\ 
 & & Var & 0.00199 & 0.00208 & 0.00206 \\ 
 & & MSE & 0.00199 & 0.00208 & 0.00206 \\ 
 \cline{3-6}
 &\multirow{3}{*}{$500$} & Bias & -0.00019 & -0.00014 & -0.00019 \\ 
 & & Var & 0.00080 & 0.00081 & 0.00081 \\ 
 & & MSE & 0.00080 & 0.00081 & 0.00081 \\ 
 \cline{3-6}
 &\multirow{3}{*}{$1000$} & Bias & -0.00006 & -0.00010 & -0.00010 \\ 
 & & Var & 0.00042 & 0.00042 & 0.00042 \\ 
  \hline
 \multirow{9}{*}{E} &  \multirow{3}{*}{$200$}& Bias & 0.00027 & -0.00001 & -0.00003 \\ 
 & & Var & 0.00195 & 0.00204 & 0.00202 \\ 
 & & MSE & 0.00195 & 0.00204 & 0.00202 \\ 
 \cline{3-6}
 &\multirow{3}{*}{$500$}  & Bias & -0.00039 & -0.00042 & -0.00044 \\ 
 & & Var & 0.00079 & 0.00080 & 0.00080 \\ 
 & & MSE & 0.00079 & 0.00080 & 0.00080 \\ 
  \cline{3-6}
 &\multirow{3}{*}{$1000$}  & Bias & -0.00005 & -0.00013 & -0.00010 \\ 
 & & Var & 0.00042 & 0.00043 & 0.00043 \\ 
 & & MSE & 0.00042 & 0.00043 & 0.00043 \\ 
   \hline
 \multirow{9}{*}{F} &  \multirow{3}{*}{$200$}& Bias & 0.00077 & 0.00102 & 0.00100 \\ 
 & & Var & 0.00199 & 0.00208 & 0.00206 \\ 
 & & MSE & 0.00199 & 0.00208 & 0.00206 \\ 
  \cline{3-6}
 &\multirow{3}{*}{$500$}  & Bias & -0.00002 & 0.00054 & 0.00047 \\ 
 & & Var & 0.00079 & 0.00080 & 0.00080 \\ 
 & & MSE & 0.00079 & 0.00080 & 0.00080 \\ 
  \cline{3-6}
 &\multirow{3}{*}{$1000$}  & Bias & 0.00007 & 0.00055 & 0.00060 \\ 
 & & Var & 0.00042 & 0.00043 & 0.00043 \\ 
 & & MSE & 0.00042 & 0.00043 & 0.00043 \\ 
\hline
\end{tabular}
}
\end{table}

\begin{table}[!ht]
\centering
\caption{
Relative biases (R.B.)  of the proposed variance estimator, coverage rates (C.R.) of the $90\%$ and $95\%$ confidence intervals for imputed estimators under kernel ridge regression imputation 
for binary responses}\label{MAR, binary, KRR}
\begin{tabular}{ccccc}
  \hline
\multirow{2}{*}{Model} & \multirow{2}{*}{Criteria}  & \multicolumn{3}{c}{Sample Size} \\
  \cline{3-5} 
    &  &   200 & 500 & 1000 \\ 
  \hline
 \multirow{3}{*}{D} & R.B. & -0.0061 & 0.0068 & -0.0392 \\ 
 & C.R. (90\%)  & 88.6\% & 90.2\% & 90.4\% \\ 
 &  C.R. (95\%)  & 94.6\% & 94.1\% & 94.3\% \\ 
 \hline
\multirow{3}{*}{E} & R.B.  & 0.0165 & 0.0222 & -0.0487 \\ 
 &  C.R. (90\%)  & 89.2\% & 89.9\% & 89.6\% \\ 
 & C.R. (95\%)   & 94.6\% & 94.7\% & 93.9\% \\  
  \hline
 \multirow{3}{*}{F} &   R.B. & -0.0062 & 0.0187 & -0.0437 \\ 
 & C.R. (90\%)   & 89.9\% & 89.7\% & 89.9\% \\ 
 & C.R. (95\%)   & 94.7\% & 94.8\% & 94.3\% \\ 
 \hline
\end{tabular}
\end{table}

%{We also compute the confidence intervals  using the asymptotic normality of the kernel ridge regression imputed estimator. The proposed variance estimator is used in computing the confidence intervals. Table 3 shows the coverage rates of the confidence intervals. The realized coverage probabilities are close to the nominal coverage probabilities, confirming the validity of the proposed interval estimator. }

\section{Application}

\begin{figure}[!ht]
    \centering  \label{CI_Fig}
            \caption{Estimated mean $\mbox{PM}_{2.5}$ concentration in December 2012 with 95\% confidence interval.}
    \label{fig:mesh1}
    \includegraphics[width=1\textwidth]{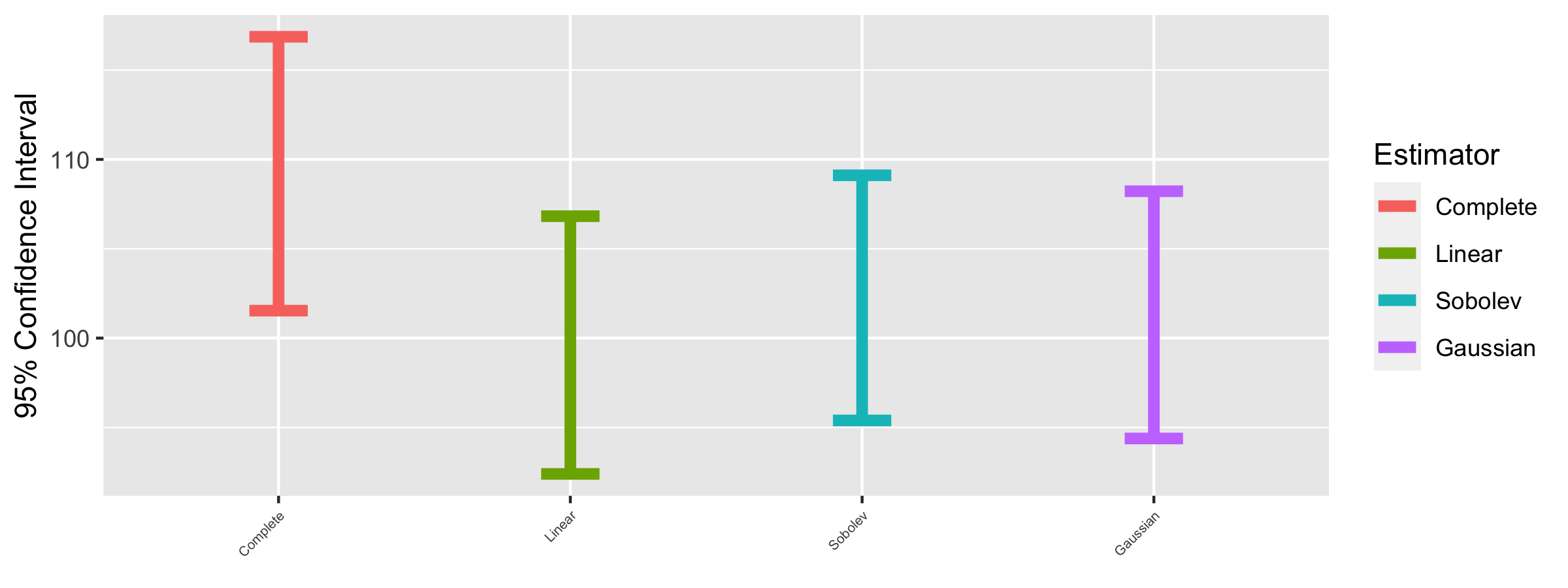}
\end{figure}

We applied the KRR with kernels of second-order Sobolev space and Gaussian kernel to study the $\mbox{PM}_{2.5}(\mu g/m^{3})$ concentration measured in Beijing, China \citep{liang2015assessing}. Hourly weather conditions: temperature, air pressure, cumulative wind speed, cumulative hours of snow and cumulative hours of rain are available from 2011 to 2015. Meanwhile, the averaged sensor response is subject to missingness. In December 2012, the missing rate of $\mbox{PM}_{2.5}$ is relatively high with missing rate $17.47\%$. We are interested in estimating the mean $\mbox{PM}_{2.5}$ in December with imputed KRR estimates. The point estimates and their 95\% confidence intervals are presented in the Table \ref{CI_Table}. The corresponding results are presented in the Figure \ref{CI_Fig}.  As a benchmark, the confidence interval computed from complete cases (Complete in Table \ref{CI_Table}) and confidence intervals for the imputed estimator under linear model (Linear) \citep{kim2009unified}  are also presented there. 

\begin{table}[!ht]
\centering
\caption{Imputed estimates (I.E.), standard error (S.E.) and $95\%$ confidence intervals (C.I.) for imputed mean $\mbox{PM}_{2.5}$ in December, 2012 under kernel ridge regression}\label{CI_Table}
\begin{tabular}{cccc}
  \hline
 Estimator & I.E. & S.E. & $95\%$ C.I. \\ 
  \hline
  Complete & 109.20 & 3.91 & (101.53, 116.87) \\ 
  Linear & 99.61 & 3.68 & (92.39, 106.83) \\ 
  Sobolev & 102.25 & 3.50 & (95.39, 109.12) \\ 
  Gaussian & 101.30 & 3.53 & (94.37, 108.22) \\ 
   \hline
\end{tabular}
\end{table}

As we can see, the performances of {KRR imputation estimators are similar} and created narrower $95\%$ confidence intervals. Furthermore,  the imputed $\mbox{PM}_{2.5}$ concentration during the missing period 
is relatively lower than  the fully observed weather conditions on average.  Therefore, if we only utilize the complete cases to estimate the  mean of $\mbox{PM}_{2.5}$, the severeness of air pollution would be over-estimated.

\section{Discussion}
 We consider kernel ridge regression  as a tool for nonparametric imputation and establish its asymptotic properties. In addition, we propose a linearized approach for variance estimation of the imputed estimator. For variance estimation, we also propose a novel approach of the maximum entropy method for  propensity score estimation.   The proposed Kernel ridge regression imputation can be used as a general tool for nonparametric imputation. By choosing different kernel functions, different  nonparametric imputation methods can be developed. The unified theory developed in this paper can cover various type of the kernel ridge regression imputation and enables us to make valid statistical inferences about the population means.

% Numerical studies confirm our theoretical results.

There are several possible extensions of the research. First, the theory can be directly applicable to other nonparametric imputation methods, such as smoothing splines \citep{claeskens2009}. Second, instead of using ridge-type penalty term, one can also consider other penalty functions such as SCAD penalty \citep{FL2001} or adaptive Lasso \citep{zou2006}. Also, the maximum entropy method for propensity score estimation should be investigated more rigorously. Such extensions will be future research topics. 

\appendix

\section*{Appendix} 
\subsection*{A. 
Proof of Theorem 1 }

\renewcommand{\theequation}{A.\arabic{equation}}
\setcounter{equation}{0}

Before we prove the main theorem, we first introduce the following lemma. 

\begin{lemma}[modified Lemma 7 in \citet{zhang2013divide}]\label{order}
Suppose Assumption [A1] and [A2] hold, for a random vector $\bz = \mathbb{E}(\bz) + \sigma \bvar$, let $\wt{\lambda} = \lambda/n$ we have
\begin{align}
 \bS_{\lambda}\bz = \mathbb{E}(\bz \mid \bx) + \mathcal{O}_{p}\left( \wt{\lambda} + \sqrt{\frac{ \gamma(\wt{\lambda})  }{n} } \right )\bone_{n},\notag
\end{align}
as long as $\mathbb{E}(\norm{z_{i}}_{\mathcal{H}})$ and $\sigma^{2}$ is bounded from above, for $i=1, \ldots, n$,  where $\bvar$ are noise vector with mean zero and bounded variance and 
 \begin{equation}\label{effective_dimension}
 \gamma(\wt{\lambda}) := \sum_{j=1}^{\infty} \frac{1}{1+\wt{\lambda}/\mu_{j}},\notag
 \end{equation}
is the effective dimension and $\{\mu_{j}\}_{j=1}^{\infty}$ are the eigenvalues of kernel $K$ used in $\hat{m}(\bx)$. 
%See Lemma \ref{Mercer} for the definition of the eigenvalues of kernel $K$
\end{lemma}
%\textcolor{red}{What is $\gamma(\wt{\lambda}) $? }

Now, to prove our main theorem, we write 
\begin{align}
\wh{\theta}_{I} &= \frac{1}{n} \sum_{i=1}^{n}\left\{ \delta_{i}y_{i} + (1-\delta_{i}) \wh{m}(\bx_{i}) \right\} \notag \\
&= \underbrace{\frac{1}{n}\sum_{i=1}^{n}m(\bx_{i})}_{:= R_{n}} + \underbrace{\frac{1}{n}\sum_{i=1}^{n}\delta_{i}\left\{   y_{i}  - m(\bx_{i}) \right\}}_{:= S_{n} }  + \underbrace{\frac{1}{n}\sum_{i=1}^{n} (1-\delta_{i})\left\{ \wh{m}(\bx_{i}) - m(\bx_{i})  \right\}}_{:=T_{n}}.
\end{align}
Therefore, as long as we show 
\begin{align}
T_{n} = \frac{1}{n} \sum_{i=1}^{n}\delta_{i}\left\{ \frac{1}{\pi(\bx_{i})} - 1  \right\}\left\{ y_{i} - m(\bx_{i})  \right\} + o_{p}(n^{-1/2}),\label{12} 
\end{align}
then the main theorem automatically holds.

To show (\ref{12}), recall that the KRR can be regarded as the following optimization problem
\begin{align}
 \wh{\balpha}_{\lambda} = \argmin_{\balpha \in \mathbb{R}^{n}} (\by - \bK\balpha)\trans \bDelta_{n} (\by - \bK \balpha) + \lambda \balpha\trans \bK \balpha.\notag 
\end{align}
Further, we have
\begin{align}
  \wh{\balpha}_{\lambda} = \left( \bDelta_{n} \bK + \lambda \bI_{n}  \right)^{-1} \bDelta_{n}\by,\notag
\end{align}
and 
\begin{align}
 \wh{\bmm} &= \bK \left(  \bDelta_{n}\bK + \lambda \bI_{n}    \right)^{-1} \bDelta_{n}\by \notag \\
 &= \bK  \left\{  \left(  \bDelta_{n} + \lambda \bK^{-1}    \right) \bK     \right\}^{-1} \bDelta_{n} \by \notag \\
 &=  \left(    \bDelta_{n} + \lambda \bK^{-1}   \right)^{-1}  \bDelta_{n}\by,\notag
\end{align}
where $\wh{\bmm} = (\wh{m}(\bx_{1}), \ldots, \wh{m}(\bx_{n}))\trans$. Let $\bS_{\lambda} =  (  \bI_{n} + \lambda \bK^{-1}   )^{-1}$, we have
\begin{align}
 \wh{\bmm} = \left( \bDelta_{n} + \lambda \bK^{-1}  \right)^{-1}\bDelta_{n}\by = \bC_{n}^{-1} \bd_{n},\notag
\end{align}
where
\begin{align}
\bC_{n} &= \bS_{\lambda}\left( \bDelta_{n} + \lambda\bK^{-1}   \right),\notag\\
\bd_{n} &= \bS_{\lambda}\bDelta_{n}\by\notag.
\end{align}

By Lemma \ref{order}, let $\wt{\lambda} = \lambda/n$, we obtain
\begin{align}
\bC_{n} &= \mathbb{E}(\bDelta_{n} \mid \bx) + \mathcal{O}_{p}\left( \wt{\lambda} + \sqrt{\frac{ \gamma(\wt{\lambda})  }{n} } \right )\bone_{n} \notag \\
&:= \bPi + \mathcal{O}_{p}\left( \wt{\lambda} + \sqrt{\frac{ \gamma(\wt{\lambda})  }{n} } \right )\bone_{n},\notag
\end{align}
where $\bPi = \mbox{diag}(\pi(\bx_{1}), \ldots, \pi(\bx_{n}))$ and 
 $\gamma(\wt{\lambda})$ is the effective dimension of kernel $K$. Similarly, we have 
\begin{align}
\bd_{n} &= \mathbb{E}(\bDelta_{n}\by \mid \bx) + \mathcal{O}_{p}\left( \wt{\lambda} + \sqrt{\frac{ \gamma(\wt{\lambda})  }{n} }\right)\bone_{n} \notag \\
&= \bPi\bmm + \mathcal{O}_{p}\left( \wt{\lambda} + \sqrt{\frac{ \gamma(\wt{\lambda})  }{n} } \right )\bone_{n}.\notag
\end{align}
Consequently, letting $a_n = \wt{\lambda} +  \sqrt{\gamma(\wt{\lambda})/n} $ and applying 
 Taylor expansion, we have
\begin{align}
 \wh{\bmm} &= \bmm + \bPi^{-1} \left(    \bd_{n} - \bC_{n}\bmm      \right) + o_{p}\left( a_n  \right )\bone_{n}\notag\\ 
 &= \bmm + \bPi^{-1} \left\{   \bS_{\lambda}\bDelta_{n}\by - \bS_{\lambda}\left( \bDelta_{n} + \lambda\bK^{-1}   \right)\bmm      \right\} \notag\\
 &\quad+ o_{p}\left( a_n  \right )\bone_{n}\notag \\
 &= \bmm + \bPi^{-1} \bS_{\lambda}\bDelta_{n}  \left(  \by - \bmm      \right) + \mathcal{O}_{p}\left(a_n  \right )\bone_{n},\notag
 \end{align}
where the last equality holds because 
\begin{align}
\bS_{\lambda} \lambda\bK^{-1}\bmm &= \bS_{\lambda}\left\{ \left( \bI_{n} + \lambda\bK^{-1} \right) - \bI_{n}  \right\}\bmm \notag \\
&= \bmm -  \bS_{\lambda} \bmm = \mathcal{O}_{p}\left( a_n  \right ). \notag 
\end{align}
%where the last equality is by Lemma \ref{order}.

Therefore, we have
 \begin{align}
T_{n} &= n^{-1} \bone\trans \left(\bI_{n} - \bDelta_{n}\right) (\wh{\bmm} - \bmm) \notag \\
&= {n}^{-1} \bone\trans \left(\bI_{n} - \bDelta_{n}\right) \bPi^{-1} \bS_{\lambda}\bDelta_{n}  \left(  \by - \bmm      \right) +  \mathcal{O}_{p}\left( a_n  \right)\notag \\
& = n^{-1} \bone\trans \left(\bI_{n} -  \bPi \right) \bPi^{-1} \bDelta_{n}  \left(  \by - \bmm      \right) +  \mathcal{O}_{p}\left( a_n  \right) \notag \\
 &= n^{-1} \bone\trans \left(\bPi^{-1} - \bI_{n}\right) \bDelta_{n}  \left(  \by - \bmm      \right) +  \mathcal{O}_{p}\left( a_n  \right)
\notag. 
\end{align}

By Corollary 5 in \citet{zhang2013divide}, for $\ell$-th order of Sobolev space, we have
\begin{align}
  \gamma(\wt{\lambda})  &= \sum_{j=1}^{\infty} \frac{1}{1 + j^{2\ell}\wt{\lambda} } \notag\\
  &\leq \wt{\lambda}^{-\frac{1}{2\ell}}  + \sum_{j >\wt{\lambda} ^{-\frac{1}{2\ell}}} \frac{1}{ 1 + j^{2\ell} \wt{\lambda}} \notag\\
  &\leq \wt{\lambda}^{-\frac{1}{2\ell}} + \wt{\lambda}^{-1} \int_{\wt{\lambda}^{-\frac{1}{2\ell}}}^{\infty}z dz \notag \\
  &= \wt{\lambda}^{-\frac{1}{2\ell}}  + \frac{1}{2\ell-1}\wt{\lambda} ^{-\frac{1}{2\ell}} \notag\\
  &= O\left(\wt{\lambda}^{-\frac{1}{2\ell}}\right).
\end{align}
Consequently, as long as $\wt{\lambda}^{-\frac{1}{2\ell}} / n = o(1)$ and $\wt{\lambda} = o(n^{-1/2})$, we have
\begin{align} 
T_{n} &=  \frac{1}{n}\bone\trans \left(\bPi^{-1} - \bI_{n}\right) \bDelta_{n}  \left(  \by - \bmm      \right) + o_{p}(n^{-1/2}).
\end{align}
One legitimate of such $\wt{\lambda}$ can be chosen as $n^{-\ell}$, i.e., $\lambda = \mathcal{O}(n^{1-\ell})$.

\subsection*{B. Computational Details }

As the objective function in \eqref{entropy_method} is convex \citep{nguyen2010}, we apply the limited-memory Broyden-Fletcher-Goldfarb-Shanno (L-BFGS) algorithm to solve 
the optimization problem with the following first order partial derivatives:
\begin{align}
  \frac{\partial U }{\partial \alpha_{0}} = & 
   \frac{1}{n_{1}}\sum_{i=1}^{n}\mathbb{I}(\delta_{i} = 0)\exp\left( \alpha_{0} + \sum_{j=1}^{n}\alpha_{j}K(x_{i}, x_{j})  \right) - 1,\notag\\
   \frac{\partial U }{\partial \alpha_{k}} = & \frac{1}{n_{1}}\sum_{i=1}^{n}\mathbb{I}(\delta_{i} = 0)k(x_{i}, x_{k})\exp\left( \alpha_{0} + \sum_{j=1}^{n}\alpha_{j}K(x_{i}, x_{j})  \right) 
   - \frac{1}{n_{0}}\sum_{i=1}^{n}K(x_{i}, x_{k}) \notag\\
   &+ 2  \tau\sum_{i=1}^{n}K(x_{i}, x_{k})\alpha_{i}, k = 1, \ldots, n. \notag
\end{align}

For tuning parameter selection $\tau$ in (\ref{18}),  we adopt a cross-validation (CV) strategy. In particular, we may firstly stratify the sample $S = \{1, \ldots, n\}$ into two strata 
	$S_{0} =\{i\in S: \delta_{i} = 0\}$  and $S_{1} =\{i\in S: \delta_{i} = 1\}$. 
Within each $S_{h}$, we make $K$ random partition $\mathcal{A}_{k}^{(h)}$ such that
\begin{align}
\begin{gathered}
     \bigcup_{k=1}^{K}\mathcal{A}_{k}^{(h)} = S_{h}, h = 0, 1\notag\\
     \mathcal{A}_{k_{1}}^{(h)} \bigcap \mathcal{A}_{k_{2}}^{(h)} =  \emptyset, k_{1} \neq k_{2}, k_{1}, k_{2} \in \{1, \ldots, K\}, \notag\\
     \Abs{\mathcal{A}_{1}}^{h} \approx  \Abs{\mathcal{A}_{2}}^{(h)} \approx \cdots \approx \Abs{\mathcal{A}_{K}}^{(h)},  h = 0, 1,\notag
\end{gathered}
\end{align}
where $|\cdot|$ is the cardinality of a specific set. For a fixed $\tau > 0$, the corresponding CV criterion is
\begin{align}\label{cv}
	\mbox{CV}(\tau) = \frac{1}{K}\sum_{k=1}^{K}\sum_{j\in \mathcal{A}_{k}} \tilde{L}(\delta_{j},  \hat{g}^{(-k)}(x_{j}, \tau) ),
\end{align}
where $\hat{g}^{(-k)}$ is the trained model with data with data points except for $\mathcal{A}_{k} = \mathcal{A}_{k}^{(0)} \cup  \mathcal{A}_{k}^{(1)}$. Regarding the loss function in \eqref{cv}, we can use
\begin{align}
	\tilde{L}(\delta; \hat{g}) = \mathbb{I}(\delta = 1, \hat{p}(x) < 0.5   ) + \mathbb{I}(\delta = 0, \hat{p}(x) > 0.5   ),\notag
\end{align}
where $ 
	\hat{p}(x) = n_{1} /\{  n_{1} + n_{0}\hat{g}(x)    \} $ 
as an estimator for $p(x) = Pr(\delta = 1 \mid x)$. As a result, we may select the tuning parameter $\tau$ which minimizes the CV criteria in \eqref{cv}.

\bibliographystyle{chicago}
\bibliography{ref}

\end{document}